\documentstyle[twocolumn,aps,epsfig,floats]{revtex}

\begin{document}     
%==============================================================================
\twocolumn[\hsize\textwidth\columnwidth\hsize\csname @twocolumnfalse\endcsname

\title{ Half-metallic antiferromagnets in double perovskites:   
       La$A$VRuO$_6$ ($A$=Ca, Sr, and Ba)}     
\author{ J. H. Park, S. K. Kwon and B. I. Min }        
\address{Department of Physics,          
         Pohang University of Science and Technology, 
         Pohang 790-784, Korea}     

\date{\today}     
\maketitle     
     
%===============================================================================
\begin{abstract}     
Based on the theoretical exploration of electronic structures,
we propose that the ordered double perovskites 
La$A$VRuO$_6$ and LaVO$_3$/$A$RuO$_3$ (001) superlattice 
($A =$ Ca, Sr and Ba) 
are strong candidates for half-metallic (HM) antiferromagnets (AFMs).
%La$A$VRuO$_6$ and LaVO$_3$/$A$RuO$_3$ have the
%100\% spin polarizations at the Fermi level but with zero
%total magnetic moments. 
We have shown that the HM-AFM nature in La$A$VRuO$_6$ is
very robust regardless of (i) divalent ion replacement at $A$-sites,
(ii) oxygen site relaxation, (iii) the inclusion of the Coulomb correlation, 
and (iv) cation disorder.  
A type of the double exchange interaction
is expected to be responsible for the half-metallicity and 
the antiferromagnetism in these systems. 
\end{abstract}     

]

\narrowtext  

%===============================================================================

%\section{Introduction}     
Since the observation of the room temperature colossal magnetoresistance (CMR)
phenomenon in Sr$_2$FeMoO$_6$ \cite{kobayashi}, 
intensive research efforts have been devoted to understanding
electronic and magnetic properties of double perovskites with   
$A_2BB'O_6$ formula unit (F.U.).
The high transition temperature $T_{\rm C}$ and the low field MR
in double perovskites suggest the high spin-polarization of conduction 
electrons and the half-metallic (HM) ground state \cite{kobayashi,Kim99}. 
In fact, the HM property is considered to be closely related to   
the CMR phenomena observed in various materials \cite{Pickett96,Youn}.    
It has also been proposed that the double perovskites
can be a suitable candidate for the half-metallic (HM) antiferromagnet (AFM) 
\cite{pickett}.  The HM-AFM is a nonmagnetic metal,
but its conduction electrons are perfectly spin-polarized. 
The HM-AFM is expected to play a vital role in the advanced spintronic devices 
that utilize the spin polarization of the conduction carriers.    
Furthermore, the success in synthesizing the ordered La$_2$CrFeO$_6$ 
as an artificial superlattice of (111) layers of LaFeO$_3$/LaCrO$_3$
stimulates research on developing new double perovskites 
with exotic properties \cite{ueda}.
The purpose of present work is to search for candidate materials
in double perovskites having the HM-AFM characteristics. 

The first HM-AFM was proposed by van Leuken and de Groot\cite{Groot}
on the basis of the Heusler compound.   
Pickett\cite{pickett} has also suggested that the double perovskite 
La$_2$VMnO$_6$ can be a promising candidate for the HM-AFM.
The local spin-density approximation (LSDA) band calculation in La$_2$VMnO$_6$
indicates that only the minority $t_{2_g}$ bands of both V and Mn 
contribute to the density of states (DOS) near the Fermi energy $E_{\rm F}$.
The valence electron configurations are Mn$^{3+}$ ($3d^4$) 
and V$^{3+}$ ($3d^2$), and the antiparallel alignment 
of the low spin Mn$^{3+}$ ($t_{2g}^3$$\uparrow$$t_{2g}^1$$\downarrow$, $S=1$) 
and the V$^{3+}$ ($t_{2g}^2$$\downarrow$, $S=1$) states yields zero 
total magnetic moment. 
The HM-AFM nature of La$_2$VMnO$_6$ has not been tested 
experimentally yet. However, the low spin state of Mn$^{3+}$ 
in the double perovskite seems to be rather unlikely.
On the other hand, it has been suggested that mixed-cation double
perovskites of $AA'BB'O_6$-type form ordered structures 
and provide good candidates for the HM-AFM \cite{pickett}.
Indeed, ordered LaBa$M$RuO$_6$ ($M =$ Mg, Zn) were synthesized 
recently\cite{hong}. However, since $M$'s are nonmagnetic elements, 
it is not a half-metal \cite{Park01}.

Motivated by the above theoretical and experimental works, 
we have explored the $AA'BB'O_6$-type double perovskites
to search for possible HM-AFMs.  
We have chosen La$A$VRuO$_6$ (L$A$VRO: $A$=Ca, Sr and Ba) as potential 
candidates. 
% which are isoelectronic to La$_2$CrRuO$_6$.   
In the ionic picture, V and Ru ion are expected to be trivalent 
V$^{3+}$ ($3d^2$) and tetravalent Ru$^{4+}$ ($4d^4$), respectively.
Then V$^{3+}$ ion has the spin moment of $2\mu_B$ in high spin state 
while Ru$^{4+}$ ion has $2\mu_B$ in low spin state, and so, if they are 
antiferromagnetically coupled, the total magnetic moment will be zero. 
Another valence configuration of V$^{4+}$ ($3d^1)$ and Ru$^{3+}$ ($4d^5$),    
albeit not so plausible, would also produce zero total magnetic moment.    
%As yet, no experimental study on synthesis 
%and physical properties of L$A$VRO has been reported.
%, while the isostructural alloys of 
%LaBa$M$RuO$_6$ ($M$=Mg,Zn,Fe,Co,Ni, and Y)    
%have been reported to have double perovskite structure \cite{hong,fern}.   
   
%===============================================================================
We have investigated electronic structures of    
mixed-cation double perovskites L$A$VRO ($A$=Ca, Sr, Ba) using both the LSDA
and the LSDA + $U$ ($U$: Coulomb correlation interaction) scheme on the basis 
of the linearized muffin-tin orbitals (LMTO) band method \cite{skkwon}.
We have considered L$A$VRO as a combined form of LaVO$_3$ and 
$A$RuO$_3$ perovskites. La$A$VRO with the antiferromagnetic coupling
of V and Ru spins corresponds to a superlattice having the layered structure 
of stacking along the [111] direction, as in La$_2$CrFeO$_6$.  
In the [111] stacking, each Ru atom has six V neighbors, 
and each V atom has six Ru neighbors.    
In reality, the fabrication of L$A$VRO film along the [111] direction   
may not be easy due to the absence of proper substrate 
and the cation disorder effect \cite{sarma}.    
Hence, in this work, we have also considered
LaVO$_3$/$A$RuO$_3$ superlattice stacking along the [001] direction 
to investigate the stable electronic and magnetic structures.    
In the [001] stacking of LaVO$_3$/$A$RuO$_3$ superlattice,
each Ru atom has four Ru and two V neighbors, 
while each V atom has four V and two Ru neighbors.   
By considering two stacking structures,
one can expect to simulate the effect of the cation disorder on
the electronic structures.
%The von Barth-Hedin form of the exchange-correlation potential is utilized,    
%and 60 $k$-points in the irreducible Brillouin zone are used 
%for the self-consistency. 
  
LaVO$_3$ is known to be a Mott insulator 
with the antiferromagnetic ordering \cite{Imada}.
Due to the $3d^2$ valency of V$^{3+}$ ion,
LaVO$_3$ would undergo a Jahn-Teller structural distortion \cite{Khaliu}. 
Also the band calculation should take into account the $3d$ electron 
Coulomb correlation because of the Mott insulating nature of LaVO$_3$
\cite{pari,Sawada}. 
Indeed, with $U=5.0$ eV and $a/c$ = 0.95 tetragonal distortion from the 
cubic symmetry, we have found
that LaVO$_3$ becomes an insulator \cite{Park02}. 
%The calculated equilibrium lattice constant of LaVO$_3$ is    
%comparable to the experimental value within one percent.
It has been reported that, with hole doping in LaVO$_3$,
the insulator to metal transition occurs: La$A$VO$_3$ ($A$=Ca, Sr)
\cite{Maiti,Miyasaka}.

%......................................................................       
\begin{figure}[t]   
\epsfig{file=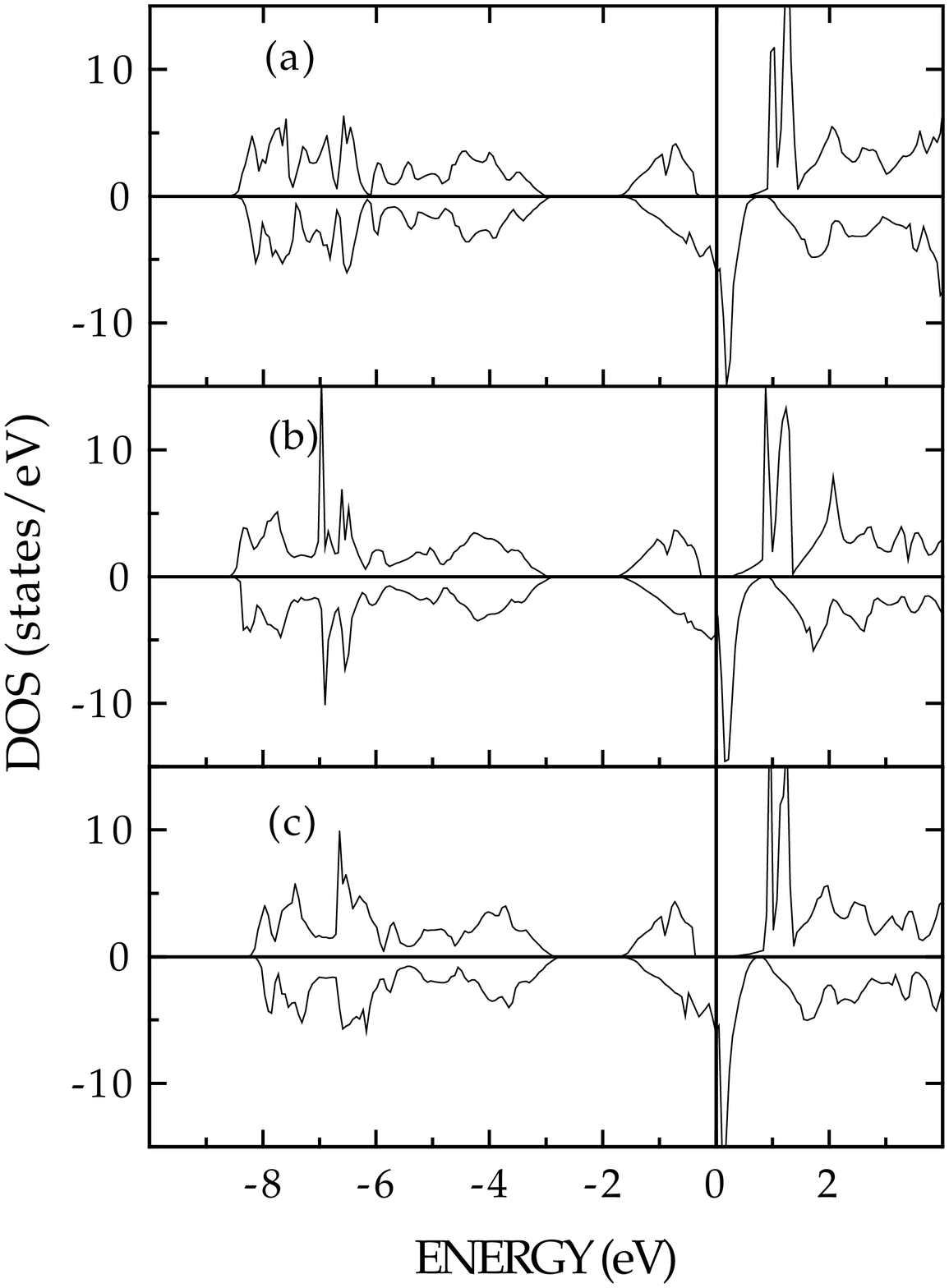,width=8.0cm}        
\caption{The total DOSs of La$A$VRuO$_6$ in the LSDA.
(a) $A =$ Ca, (b) $A =$ Sr and (c) $A =$ Ba.}     
\label{dos111}         
\end{figure}         
%......................................................................       
Both CaRuO$_3$ and SrRuO$_3$ exhibit metallic conductivity. 
CaRuO$_3$ has an orthorhombically distorted perovskite structure 
with GdFeO$_3$-type ($Pbnm$) \cite{catchen}.
SrRuO$_3$ also has an orthorhombic crystal structure \cite{catchen}.
For the magnetic property of CaRuO$_3$, several conflicting results 
are reported \cite{callaghan,longo,gibb}. 
According to the most recent report, CaRuO$_3$ is considered 
to be a paramagnetic metal\cite{cao}.  
On the other hand, SrRuO$_3$ is a ferromagnet with $T_{\rm C} \sim 160$ K 
and the magnetic moment $M=1.4\pm0.4\mu{_B}$ per Ru atom 
\cite{callaghan,longo}.
BaRuO$_3$ is chemically related and isoelectronic to CaRuO$_3$ and SrRuO$_3$, 
but its structure and electronic properties are very different.  
The crystal structure of BaRuO$_3$ belongs 
to the hexagonal perovskite family  
with various crystallographic forms, $4R$, $6R$, and $9R$, 
according to the stacking sequence. 
The $4R$-BaRuO$_3$ shows a metallic resistivity behavior 
down to low temperature, 
while the $9R$-BaRuO$_3$ shows a crossover to a more 
resistive state \cite{rijssen}.
%The valence band photoemission revealed that BaRuO$_3$ is  
%a metallic material with a well-defined cutoff at $\rm E_F \cite{gulino}. 
The magnetic states are paramagnetic for all 
BaRuO$_3$ \cite{rijssen,gulino}. 
%-----------------------------------------------
%-----------------------------------------------

The LSDA equilibrium lattice constant of LaSrVRuO$_6$ (LSrVRO)
turns out to be close to the average value of the lattice constants   
of its parent materials, LaVO$_3$ and SrRuO$_3$ \cite{Delh}.
The ferrimagnetic ground state is lower in energy than
the paramagnetic state by 0.24 eV/F.U. for LSrVRO. 
Likewise, both LaCaVRuO$_6$ (LCaVRO) and
LaBaVRuO$_6$ (LBaVRO) have the ferrimagnetic ground states too 
(see Table \ref{table1}).
The total DOSs of L$A$VRO ($A =$ Ca,Sr, and Ba) are presented 
in Fig. \ref{dos111} which shows the HM-AFM nature of all three systems.
As is well known, $A$-site atoms in $AB$O$_3$ perovskite act as carrier     
reservoir and volume conserver.
The interaction between $A$ and neighboring atoms is 
so weak that many $AB$O$_3$'s even with different $A$ shares 
similar electronic properties. 
Indeed, as seen in Fig. \ref{dos111}, the DOSs are similar for all
L$A$VRO, even if the $A$-site atom is varied.
Most interestingly, the antiparallel alignment of Ru and V magnetic
moments yields zero total magnetic moment for all cases,
by counting together the polarized magnetic moments of La, $A$, and O atoms.
The local magnetic moments of Ru and V in L$A$VRO are provided 
in Table \ref{table1}. 

%-------------------------------------------------------------------------------
\begin{table}[b]    
\caption{\label{table1} The equilibrium lattice constants ($a$) 
and the total energy difference ($\Delta E$ in eV/F.U)
between the ferrimagnetic and paramagnetic states.
$M_{\rm tot}$ is the total magnetic moment in $\mu_B$ per F.U.,
and $M_{\rm V}$ ($M_{\rm Ru}$) are local magnetic moments
at V (Ru) sites.
}
\begin{tabular}{lccc}   
                    & LaCaVRuO$_6$ & LaSrVRuO$_6$ & LaBaVRuO$_6$ \\ \hline    
$a$ (\AA)           &  7.767       &  7.867       &     7.965    \\
$\Delta E$          & $-0.10$~~~~  & $-0.24$~~~~  & $-0.12$~~~~  \\ 
$M_{\rm tot}$       &  0.00~~      &  0.00~~      &  0.00~~      \\
$M_{\rm V}$         & $-1.22$~~~~  & $-1.26$~~~~  & $-1.37$~~~~  \\
$M_{\rm Ru}$        &  1.01~~      &  1.05~~      &  1.10~~      
\end{tabular}      
\end{table}
%-------------------------------------------------------------------------------

%......................................................................       
\begin{figure}[t]
\epsfig{file=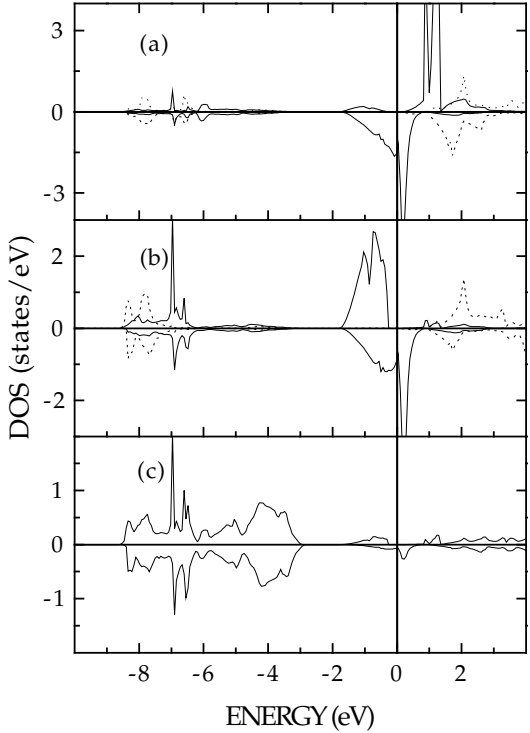,width=8.0cm}        
\caption{The orbital projected DOS of LaSrVRuO$_6$ in the LSDA.
Solid and dotted lines in (a) and (b) denote $t_{2g}$ and $e_g$ states,
respectively.
(a) V $3d$, (b) Ru $4d$ and (c) O $2p$. }     
\label{pdos111}         
\end{figure}         
%......................................................................       

For LSrVRO, as shown in Fig. \ref{pdos111}, 
the energy gap of $\sim$ 0.8 eV opens 
between the occupied Ru $t_{2g}$ and the empty V $t_{2g}$ states 
in the spin-up bands.
On the other hand, the spin-down bands are metallic 
and composed of $t_{2g}$ states of Ru and V. 
The DOSs of two $t_{2g}$ states near $E_{\rm F}$ are similar 
in the weight and the shape. 
Occupied $t_{2g}$ states in Fig. \ref{pdos111} seem to reflect the 
nominal valences of V$^{3+}$ ($d^2$) and Ru$^{4+}$ ($d^4$), respectively.
In fact, at Ru site, the spin-up bands are occupied 
by nearly three $t_{2g}$ electrons,   
while the spin-down bands by about 1.8 $t_{2g}$ electrons.
At V site, the spin-down bands are occupied by about 1.6 $t_{2g}$ electrons,
while the spin-up bands are nearly empty.   
Hence, the electron configurations become V$^{3.4+}$ ($3d^{1.6}$)
and Ru$^{3.2+}$ ($4d^{4.8}$).
It is thus likely that two valence states of V$^{3+}$/Ru$^{4+}$ 
and V$^{4+}$/Ru$^{3+}$ are nearly degenerate 
to produce a type of the double exchange (DE) interaction \cite{Zener}.
That is, the hopping of itinerant $t_{2g}$ spin-down electrons 
between Ru-V sites yields the kinetic energy gain so as to 
induce the ferrimagnetism between Ru-V spins. 
This mechanism  explains both 
the half-metallicity and the ferrimagnetism in LSrVRO.
Note that the spins of itinerant carriers are antiparallel to the localized 
Ru spins ($t_{2g}^3$$\uparrow$), satisfying the Hund rule in the low
spin state, as in Sr$_2$FeMoO$_6$ and Sr$_2$FeReO$_6$ \cite{sleight,Kang}.      
The $e_g$ states of Ru and V are located about 2 eV above $E_{\rm F}$, 
and the O $2p$ states are located between 3 eV and 8 eV  below $E_{\rm F}$.   

\begin{figure}[t]
\epsfig{file=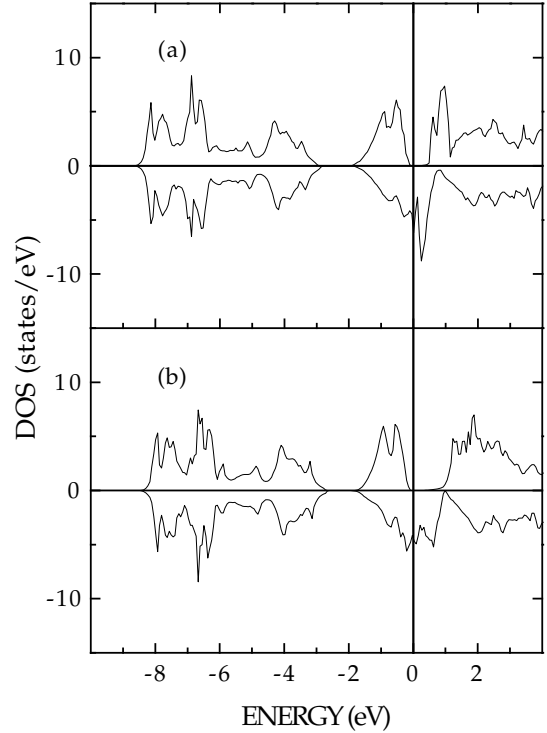,width=8.0cm}        
\caption{The total DOS of LaSrVRuO$_6$ with the oxygen site relaxation of
$d$(V-O)/$d$(Ru-O) = 0.9 (a) in the LSDA  and (b) in the LSDA + $U$ 
($U =$ 3.0 eV for V $3d$ electrons).
}     
\label{dosU}         
\end{figure}

We have checked the effect of oxygen site relaxation 
which is an important factor in double perovskites \cite{hong,moritomo}.   
In the case of LaSr$M$RuO$_6$ ($M =$ Zn, Mg), the oxygen site is reported to
be relaxed with $d$(Ru-O)/$d$(M-O) = 0.96 \cite{hong}.    
This value is similar to those observed in other double perovskites 
\cite{kobayashi,moritomo}. 
The inclusion of oxygen site relaxation with $d$(Ru-O)/$d$(V-O) = 0.96 
produces essentially the same electronic structure as 
in the case of the ideal structure.   
The HM nature is retained and the total 
magnetic moment is zero. In addition, we have also considered the case of
opposite oxygen site relaxation of $d$(V-O)/$d$(Ru-O) = 0.9 
(see Fig. \ref{dosU}(a)). In this case,
the magnetic moments of both V and Ru are reduced ($M_{\rm V}=-0.65$
and $M_{\rm Ru}=0.55 \mu_B$), but the HM-AFM nature is still retained.

%......................................................................       
%......................................................................       
\begin{figure}[t]   
\epsfig{file=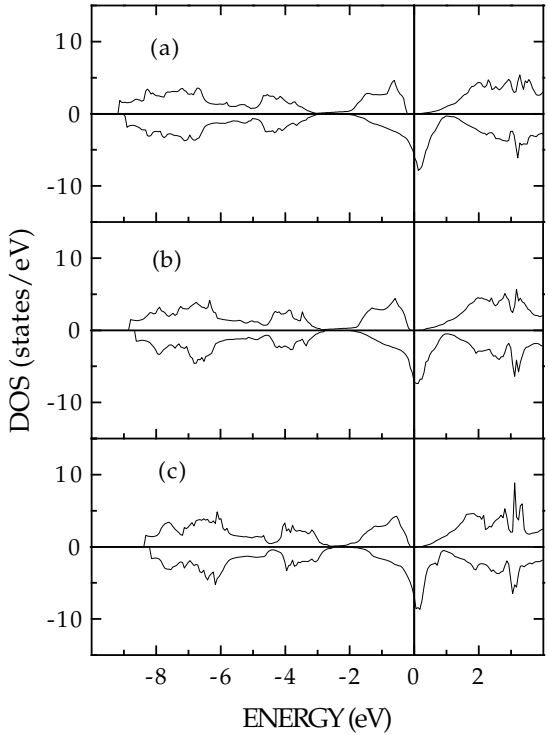,width=8.0cm}        
\caption{The total DOSs of LaVO$_3$/$A$RuO$_3$ 
in the LSDA + $U$ ($U = 2.0$ eV for V $3d$ and $U = 1.0$ eV 
for Ru $4d$ electrons). (a) $A =$ Ca, (b) $A =$ Sr and (c) $A =$ Ba.    
}     
\label{dos001}         
\end{figure}         
%......................................................................       

Since the $d$ bandwidths of Ru and V are relatively wide,   
one expects that the correlation effect in $d$ bands of L$A$VRO would be minor. 
However, as described above, LaVO$_3$ is known to be a Mott insulator. 
Hence, it might be necessary to examine the correlation 
effect of V $3d$ electrons in the electronic structure of L$A$VRO.
Figure \ref{dosU}(b) provides the DOS for L$Sr$VRO 
using the LSDA + $U$ method with $U=3.0$ eV and $J = 0.87$ eV 
for V $3d$ electrons.
Here $d$(V-O)/$d$(Ru-O) = 0.9 is 
assumed as in Fig. \ref{dosU}(a). The LSDA + $U$
yields no qualitative difference in electronic and magnetic properties 
from those of the LSDA. Magnetic moments of V and Ru are
somewhat enhanced, $-1.01$ and $0.81 \mu_B$, respectively, but
the HM-AFM nature is maintained in this case too.

In Fig. \ref{dos001} we have plotted the DOSs of LaVO$_3$/$A$RuO$_3$ [001] 
superlattice in the LSDA + $U$ calculation.
We have employed $U = 2.0$ eV and $J = 0.87$ eV for V $3d$ and $U = 1.0$ eV 
and $J = 0.87$ eV for Ru $4d$ electrons, respectively.    
Interestingly, still in [001] superlattice, 
the HM-AFM is the ground state. 
It is seen that the DOSs are a bit broader  
than those of double perovskite L$A$VRO. 
It is because extra hybridizations
between the same ions (La-La,V-V,$A$-$A$,Ru-Ru) 
exist in the LaVO$_3$/$A$RuO$_3$ [001] superlattice.   
This result suggests that the HM-AFM nature of L$A$VRO is 
maintained even though there may be some cation disorder in 
the layers \cite{Diso}.   
Therefore, LaVO$_3$/$A$RuO$_3$ [001] superlattice film would
be a very promising candidate for the HM-AFM, 
provided it can be synthesized successfully.

Table \ref{table1} indicates that 
the LSrVRO out of three candidates is the most promising, 
because it has the largest total energy difference 
between the ferrimagnetic state and the paramagnetic state.
It was reported that the epitaxial thin films of parent materials, 
SrRuO$_3$ and LaVO$_3$, can be grown on LaAlO$_3$ substrates \cite{Lu,Choi}.  
Further, SrRuO$_3$ and LaVO$_3$ are isostructural 
and the lattice constants are quite close.
Once LSrVRO is synthesized, 
one thing to be checked is the Curie temperature $T_{\rm C}$.
Note that $T_{\rm C}$ of SrRuO$_3$ is $\sim 165$ K 
and the N\'{e}el temperature $T_{\rm N}$ of LaVO$_3$ is $\sim 140$ K,
which may give rise to $T_{\rm C}$ of LSrVRO at most $\sim 165$ K.
Rather, it is our expectation that $T_{\rm C}$ of LSrVRO would be much 
higher as in Sr$_2$FeMoO$_6$ due to the DE mechanism 
in which $T_{\rm C}$ is proportional to the band width of itinerant carriers. 
We would like to note that the band width of the spin-down $t_{2g}$ electrons 
in LSrVRO is as large as 2 eV. 

In conclusion, we have shown that the double perovskite L$A$VRO 
($A =$ Ca, Sr, and Ba) are strong candidates for the HM-AFMs.  
The HM-AFM nature in L$A$VRO is very robust regardless of various factors:   
$A$-site ions (Ca, Sr, and Ba), the oxygen site relaxation, 
the Coulomb correlation effect, and the cation disorder.   
We have also shown that the LaVO$_3$/$A$RuO$_3$ [001] superlattice 
would be a HM-AFM. A type of the double exchange interaction is operative 
in these systems to yield the half-metallicity and the antiferromagnetism.  
LSrVRO is expected to be the most promising candidate for a HM-AFM.
It is thus very desirable to test experimentally
the HM-AFM nature of the double perovskite LSrVRO 
and the LaVO$_3$/SrRuO$_3$ [001] superlattice film.

Acknowledgements$-$   
This work was supported by the KOSEF through the eSSC at POSTECH   
and in part by the BK21 Project.   
     
%===============================================================================
     
%===============================================================================

\begin{references}
\bibitem{Park01} It was reported that Ru has large magnetic moment 
        ($\sim 4\mu_B$), but our LSDA calculation yields 
         smaller moment of $2.0\mu_B$.
\bibitem{Delh} The enthalpy of formation ($\Delta H$) is obtained to be
negative: $\Delta H= -0.70$ eV for LSrVRO.
\end{references}
\end{document}